\documentclass[conference]{IEEEtran}
\IEEEoverridecommandlockouts

\usepackage{cite}
\usepackage{amsmath,amssymb,amsfonts,graphicx,subcaption,hyperref,siunitx}
\usepackage{graphicx}
\usepackage{textcomp}
\usepackage{xcolor}
\def\BibTeX{{\rm B\kern-.05em{\sc i\kern-.025em b}\kern-.08em
    T\kern-.1667em\lower.7ex\hbox{E}\kern-.125emX}}
\begin{document}

\title{Revising Second Order Terms in Deep Animation Video Coding
\thanks{The authors gratefully acknowledge the scientific support and HPC resources provided by the Erlangen National High Performance Computing Center (NHR@FAU) of the Friedrich-Alexander-Universität Erlangen-Nürnberg (FAU). The hardware is funded by the German Research Foundation (DFG)}
}

\author{\IEEEauthorblockN{Konstantin Schmidt}
\IEEEauthorblockA{\textit{Fraunhofer IIS} \\
Am Wolfsmantel 33, 91058 Erlangen, Germany \\
konstantin.schmidt@iis.fraunhofer.de}
\and
\IEEEauthorblockN{Thomas Richter}
\IEEEauthorblockA{\textit{Fraunhofer IIS} \\
Am Wolfsmantel 33, 91058 Erlangen, Germany \\
thomas.richter@iis.fraunhofer.de}
}

\maketitle

\begin{abstract}
First Order Motion Model is a generative model that animates human heads based on very little motion information derived from keypoints. It is a promising solution for video communication because first it operates at very low bitrate and second its computational complexity is moderate compared to other learning based video codecs. However, it has strong limitations by design. Since it generates facial animations by warping source-images, it fails to recreate videos with strong head movements. 
This works concentrates on one specific kind of head movements, namely head rotations. We show that replacing the Jacobian transformations in FOMM by a global rotation helps the system to perform better on items with head-rotations while saving 40\% to 80\% of bitrate on P-frames.
Moreover, we apply state-of-the-art normalization techniques to the discriminator to stabilize the adversarial training which is essential for generating visually appealing videos.
We evaluate the performance by the learned metics LPIPS and DISTS to show the success our optimizations.
\end{abstract}


\section{Introduction}
\label{sec:intro}
Recently there has been a lot of progress in the development of learning based video codecs for communication applications. These codecs are able to outperform classical engineered codec standards like e.g. H.264, H.265 or H.266 (AVC \cite{AVC}, HEVC \cite{HEVC}, VVC \cite{VVC}) by quite a margin \cite{9941493,10655044,10204171}. Among these learning based codecs two main paths of development can be identified. First are codecs that mimic the classical design of video codecs based on motion estimation and residual coding. These codecs use deep networks to replace the hand-engineered predictors and residual coders. The main drawback of this design is the huge computational complexity caused by the decoder running in the encoder. Such codecs are able to generate videos at bitrates more than 30 \% lower than H.266 \cite{10655044}.
Second are codecs based on First Order Motion Model (FOMM) \cite{FOMM,KONUKO,OSFV,9522751,9859867,Agarwal_2022_BMVC,Chen2023InteractiveFV,9455985,9810732,9897729,10181664,10030232,10118851,10378582,9949138,chen2020comprisesgoodtalkingheadvideo} that are able to generate appealing videos at moderate computational complexity at bitrates of 3 kbps or even lower. FOMM has less than 60 million parameters and 55 G-MACS complexity per P-frame.
These codecs are suited for communication applications only (coding of human faces), while the former are able to code general content. Such codecs use a single image (denoted here as I-frame) that is animated by as low as 10 keypoints (KPs) that are transmitted per video-frame to be coded (denoted here as P-frame).
While sometimes being able to generate almost artifact-free videos at bitrates below 3 kbps, such codecs have major limitations by design.
One such limitation is strong head rotations on the roll-axis of P-frames, since these items do not occur very often in the training set.
As a solution to this problem, we propose to amend or replace the Jacobian transformation of the KPs by a single rotation parameter. We show that this improves the quality of generated videos while also reducing the bitrate of P-frames.

The second optimization we propose is targeting the adversarial loss and could be applied to any generative facial animation codec. The basic principle of FOMM is to generate videos by warping I-frames based on some warping information coded in the bitstream.
Often the P-frame to generate contains elements that are not present in the I-frame. In such cases the warping fails to generate meaningful content and the model relies on an adversarial loss to hallucinate these parts. The adversarial loss is known to be very unstable and as a result often only a small amount is added to the overall loss. We propose two normalization techniques to stabilize the discriminator which allows for higher amount adversarial loss and thus better image quality. First, to our best knowledge we are the first to use Gradient Normalization \cite{GRAD_NORM} to stabilize the discriminator in generative video coding. Second, inspired by self-supervised learning, we let the discriminator learn facial landmarks while training. 
In summary our contributions are:
\begin{itemize}
\item Optimized linear transformation of local patches that form the output frame.
\item Reduced P-frame bitrate compared to the original FOMM.
\item A more stable adversarial loss that results in higher image fidelity.
\end{itemize}

\section{Optimizations of the Transformations for Warping I-frames}
\begin{figure}[t]
    \centering
    \begin{subfigure}[t]{.45\columnwidth}
        \centering
        \includegraphics[width=0.7\columnwidth]{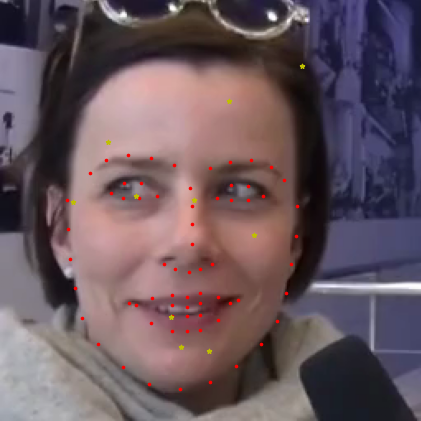}
        \caption{Landmarks (red) and Keypoints (KP).}
        \label{fig:landmarks}
    \end{subfigure}
    \begin{subfigure}[t]{.45\columnwidth}
        \centering
        \includegraphics[width=0.5\columnwidth]{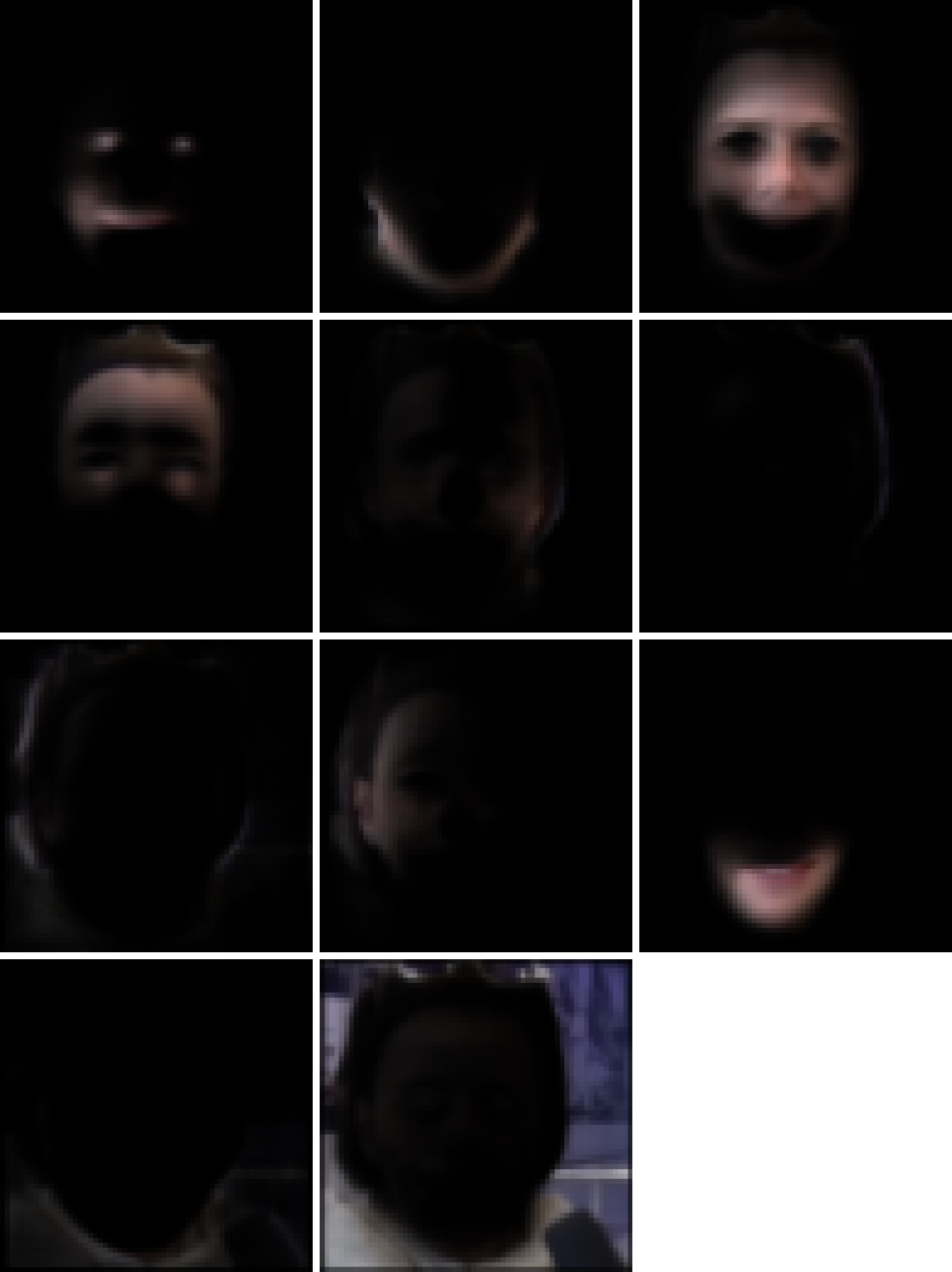}
        \caption{Warped patches with one additional patch for the background (bottom center). }
        \label{fig:features}
    \end{subfigure}
\end{figure}

\begin{figure*}[t]
    \centering
    \includegraphics[width=0.7\textwidth]{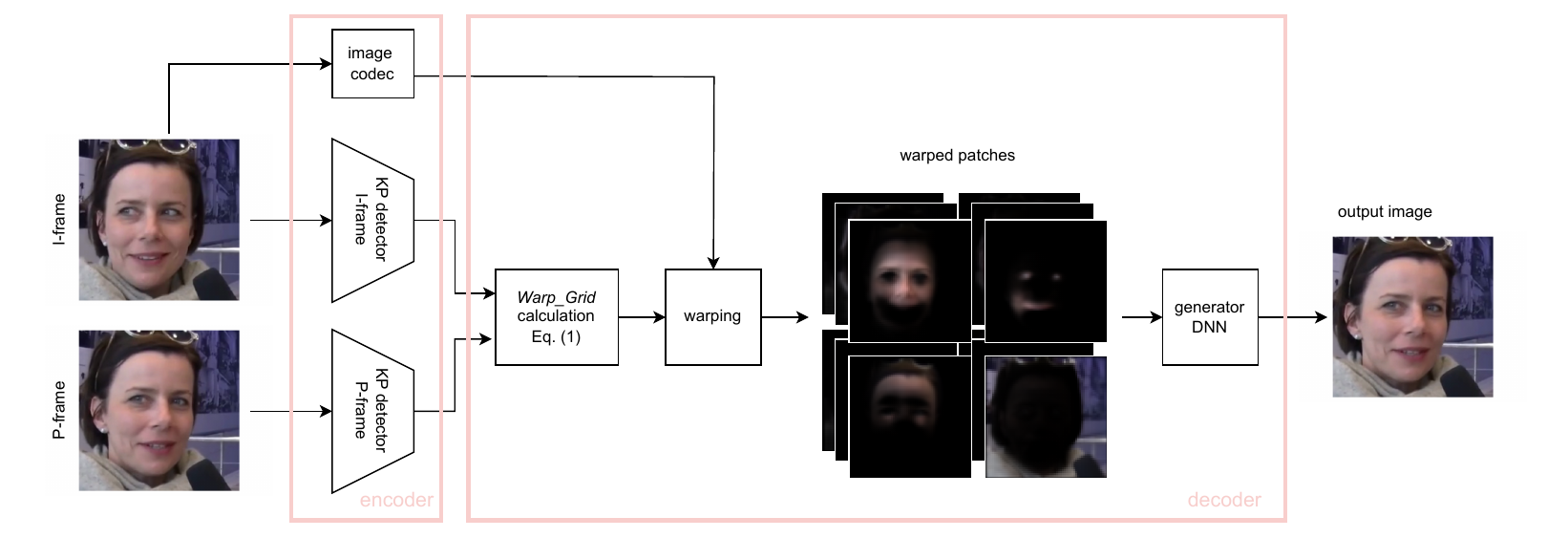}
    \caption{Block diagram of encoder and decoder in red boxes. Input are I-frame and P-frame on the left, output is the generated P-frame on the right. DNNs are depicted with green outlines.}
    \label{fig:overview}
\end{figure*}

FOMM generates animated P-frames by warping I-frame areas around so called keypoints (KP). These KPs are outputted by a DNN per video from (see Fig. \ref{fig:overview}). They are similar to facial landmarks (which identify facial elements like eyes, nose, etc.) but are learned unsupervisedly. Fig. \ref{fig:landmarks} depicts the difference between KPs and landmarks on a video-frame of an item of the training set. The 68 landmarks generated with \cite{DLIB_LANDMARKS} are plotted in red and the KPs are plotted as yellow asterisks. These KPs, together with also unsupervisedly learned local linear transformations matrices called Jacobians (JACs) form the bitstream of the P-frames. 

The warping operation used here is based on Jaderberg et. al. \cite{GRID_SAMPLE} where the warped output is calculated by the per-pixel displacement information given in the matrix $WG$ (please see \cite{GRID_SAMPLE} for more details on this). This matrix is calculated as:
\begin{equation}\label{eq:warp}
    WG = (WG_{neutral} - KP_{p} ) \cdot JAC_{i} \cdot JAC_{p}^{-1} + KP_{i},
\end{equation}
based on the unitary warping grid $WG_{neutral}$. 
$JAC_{i}$ and $JAC_{p}$ are local 2x2 linear transformation matrices of I-frame and P-frame that, together with the KPs allow for affine (i.e. first order) transformations of local patches. 
Fig. \ref{fig:features} (right) depicts such patches. Each patch contains a mutually exclusive part of the input image that is warped and merged by a generator DNN (Fig. \ref{fig:overview}) to form the output image. Please note that Eq. \ref{eq:warp} is performed for each of the 10 KPs.
After inspecting these local matrix-transformations, we conclude that:
\begin{itemize}
\item Rotations mostly follow global head rotations on the jaw-axis.
\item Scalings mostly follow global head scaling.
\item Transformations like shearing often appear random.
\item Calculating inverse $JAC$s as in \ref{eq:warp} is often unstable.
\end{itemize}
Especially during training an unstable matrix inversion is suboptimal for robust gradient flow.
This motivates us to replace the local transformation matrices $JAC_{k}$ for KP $k$ by a single global rotation and a scaling:
\begin{equation}
    JAC_{k} := R \cdot scf = 
\begin{bmatrix}
\cos \phi & - \sin \phi \\
\sin \phi & \cos \phi
\end{bmatrix}
\cdot scf,
\end{equation}
or a combination of a single global rotation, scaling and shearing matrices:
\begin{equation}\label{eq:shear}
    \begin{aligned}
    JAC_{k} & := R \cdot SHR_{k} \cdot scf \\
        & =
\begin{bmatrix}
\cos \phi & - \sin \phi \\
\sin \phi & \cos \phi
\end{bmatrix}
\cdot
\begin{bmatrix}
1+\lambda \mu_{k} & \lambda_{k} \\
\mu_{k} & 1
\end{bmatrix}    
\cdot scf.
\end{aligned}
\end{equation}
The rotation matrix 
$R=\big(\begin{smallmatrix}
 \cos \phi & - \sin \phi \\
 \sin \phi & \cos \phi
\end{smallmatrix}\big)$
has a single parameter $\phi$ that is given by the KP-detection DNN (Fig. \ref{fig:overview}) and learned by a supervised loss from a head-pose estimation loss. 
We use the model in \cite{HOPENET} as head-pose estimation network which gives a robust estimation of jaw, pitch and roll axis of the head. Here, we only use the roll axis which directly gives the parameter $\phi$. 
The inverse of such a rotation matrix is simply its transpose and is easy to calculate. As loss for learning $\phi$ we simply use an L-1 distance between $\phi$ and the head-pose estimation network output.
Using a head-pose estimation loss was first done in \cite{OSFV} but applied to a different architecture.

Since the KPs linearly scale with the size of the head in the video, we hypothesize that the scaling operation of a general 2x2 matrix can be replaced by a single scale factor $scf$ deduced from the already trasmited KPs by linear regression at decoder side:
\begin{equation}
scf = \frac{\sum_{k} (KP_{i,k} - \overline{KP_{i,k}}) * (KP_{p,k} - \overline{KP_{p,k}})}{\sum_{k} (KP_{i,k} - \overline{KP_{i,k}})^{2}},
\end{equation}
where $KP_{i,k}$ and $KP_{p,k}$ are the $k$-th KPs of the I-frame and the P-fames respectively, $\overline{KP_{i}}$ and $\overline{KP_{p}}$ are the mean of the KPs.

Replacing the per-keypoint 2x2 matrix transformations with these 2 global transformations already works well and will be evaluated as a first low-bitrate system in Sec. \ref{sec:eval}. However, sometimes the performance can be improved by a shearing operation in Eq. \ref{eq:shear} that can't be performed by the afore mentioned rotation and scaling. Such a shearing matrix contains two learnable parameters per KP. The inverse of a shearing matrix is easy to calculate:
\begin{equation}\label{eq:shear_inverse}
    SHR^{-1} = 
\begin{bmatrix}
1 & - \lambda \\
- \mu & 1+\lambda \mu
\end{bmatrix}    
\end{equation}
\label{sec:rotation}

\section{Optimizations of the Adversarial Training}
\label{sec:d_tunings}
As mentioned before, parts of the generated P-frames can be deduced from the I-frame and the warping information transmitted in the bitstream, while other parts need to be hallucinated by a generative model. The generative model used here is a generative adversarial network (GAN) which allows for good performance at moderate model size and computational complexity compared to e.g. diffusion models. However, the unstable training process remains a challenging problem and often the generator tends to fool the discriminator before learning to generate realistic images. This is usually caused by the sharp gradient space of the discriminator, which causes mode collapse in the training process of the generator. A promising solution to this problem is Gradient Normalization (GN) \cite{GRAD_NORM} which is a model-wise, non-sampling-based, and non-hard normalization of the discriminator function. A discriminator normalized with GN has increased capacity while still being Lipschitz-contrained, which ultimately results in higher fidelity of the generated content.

\begin{figure}[h]
    \centering
    \includegraphics[width=0.8\columnwidth]{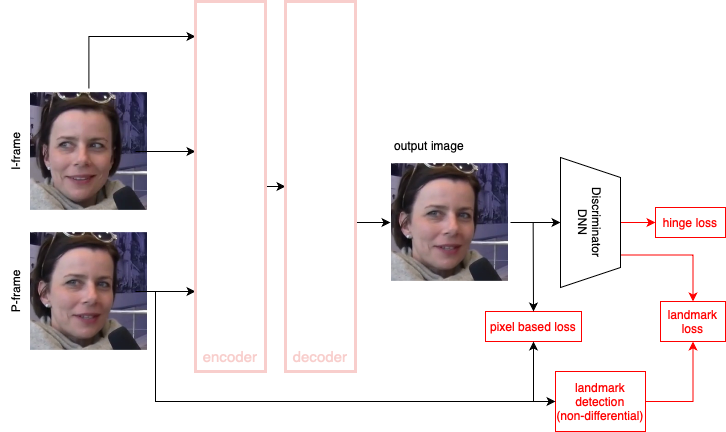}
    \caption{Block diagram of the proposed discriminator optimization. Red blocks are needed to calculate the discriminator loss.}
    \label{fig:loss}
\end{figure}

Several other systems that are based on FOMM propose to use a landmark loss as additional loss during training \cite{10378582}. This landmark loss assures that the generated images have the same facial landmarks as the original image. Here, we propose to move the landmark loss as an additional self-supervision loss after the discriminator for two reasons: First, the predominant method of landmark estimation relies on \cite{DLIB_LANDMARKS}, which is a non-differentiable model and can't directly be used as loss. Second, according to \cite{GAN_OPT_OVERVIEW,SELF_SUPERV} a self-supervision loss helps to mitigate overfitting, improves the stability and generalizability of discriminator and avoids discriminator forgetting.
The loss calculation is shown by red paths in Fig. \ref{fig:loss}.
To our best knowledge we are the first successfully applying GN and self-supervision by landmarks in generative models for video coding.

\section{Evaluation}
\label{sec:eval}
Before presenting the results, we discuss the selection of metrics we use and what data has been used for training and evaluation. 
It is known that metrics based on pixel distances like PSNR and SSIM are not able to predict the perceived quality of content created by generative models.
Among the best performing metrics to evaluate the quality of generated videos are DNN based metrics like LPISPS \cite{LPIPS} and DISTS \cite{DISTS}. These metrics estimate the quality of each video-frame and finally calculate an average over all frames. 
They are based on the hypothesis that features extracted from image classification networks are also able to estimate human perceived quality. 
According to the authors DISTS correlates better with human ratings.
PSNR and SSIM values of the presented systems are given in Tab. \ref{tab:lpips_etc} for completeness only.
To evaluate the diversity of the generated images we use Fréchet inception distance (FID) which compares the distribution of generated images with the distribution of a set of real images. A high FID score can be used to monitor mode collapse in adversarial training.

The dataset used for training is the Vox-Celeb2 dataset \cite{VoxCeleb2} which contains videos at a resolution of 256×256 pixel.
We used 32927 items for training and 6174 items for testing, with the test items not being part of the training set. The generated videos contain 90 frames, with only the first frame being sent as I-frame.
Opposed to the reference settings in \cite{FOMM} we use the AdamW optimizer \cite{ADAMW} with the parameters ($\beta_{1}=0.9$
and $\beta_{2}=0.999$) and a learning rate of $2e^{-4}$.
The training is conducted for 480 epochs on NVIDIA H100 GPU with batch size of 68 \cite{HPC}.

We also compare our system to the one presented by Wang et. al. \cite{OSFV}. Since there is no official implementation available we use the implementation from \cite{OSFV_git}. Their system is labeled as ``OSFV'' in the plots.

\begin{figure}
    \centering
    \includegraphics[width=0.9\columnwidth]{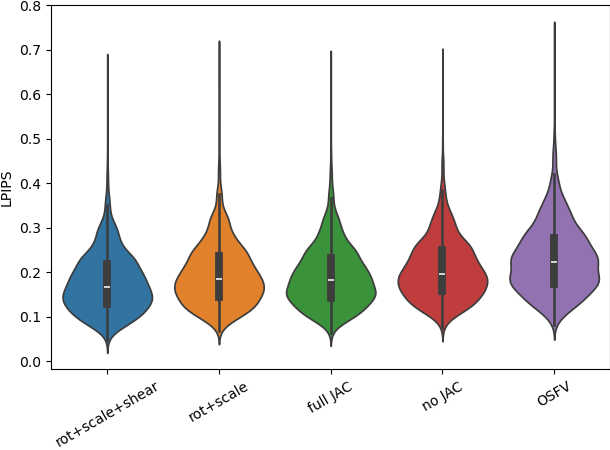}
    \caption{LPIPS on test set. Lower means better performance.}
    \label{fig:lpips}
\end{figure}

\begin{figure}
    \centering
    \includegraphics[width=0.9\columnwidth]{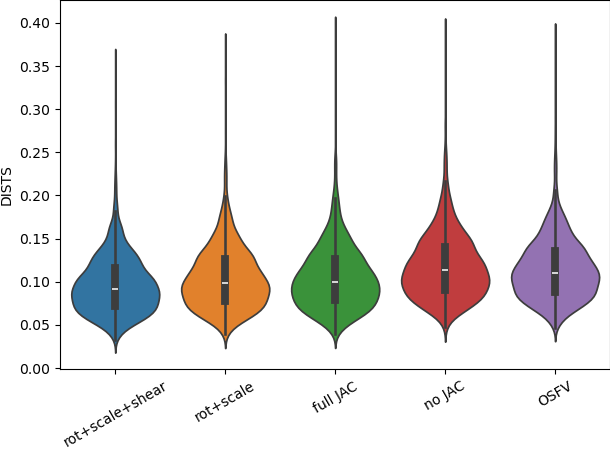}
    \caption{DISTS on test set. Lower means better performance.}
    \label{fig:dists}
\end{figure}

\begin{figure}
    \centering
    \includegraphics[width=0.9\columnwidth]{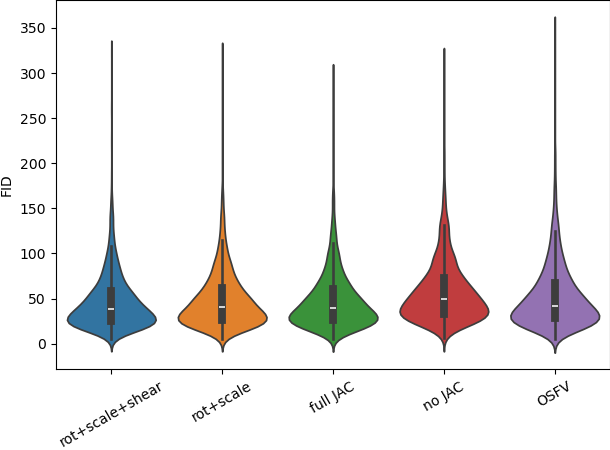}
    \caption{Fréchet inception distance (FID) on test set. Lower means better performance.}
    \label{fig:fid}
\end{figure}

The results are given in Tab. \ref{tab:lpips_etc} and Fig. \ref{fig:lpips},\ref{fig:dists},\ref{fig:fid} as distribution over all test items. The proposed rotation, scaling and shearing (\textit{rot+scale+shear}) achieving the best results in all metrics. Not sending any Jacobians at all (\textit{no JAC}) has the strongest negative impact on the quality. Sending only a global rotation (\textit{rot.+scal.}) already brings the quality close to the full Jacobian reference (\textit{full JAC}). 
It has to be emphasized that the presented systems have much lower elements in the bitstream. 
In addition to the 10 KPs that are present in all systems, FOMM needs 4 additional parameters per KP. The presented system with rotation and scaling (rot. + scale) has only one global rotation while the presented system with rotation, scaling and shearing has 2 additional parameters per KP. Estimated bitrates are also given in Tab. \ref{tab:lpips_etc}.
\begin{table}[h]
\begin{center}
\begin{tabular}{ r | l l l l l}
        & \begin{tabular}[c]{@{}c@{}}rot.+scale\\+shear\end{tabular} &rot.+scal.&full JAC&no JAC&OSFV\\ 
\hline
LPIPS $\downarrow$ & \textbf{0.179} & 0.195 & 0.192 & 0.207 & 0.231 \\ 
DISTS $\downarrow$ & \textbf{0.099} & 0.105 & 0.105 & 0.119 & 0.115 \\  
FID $\downarrow$ & \textbf{47.70}  & 48.95  & 47.87  & 57.77 & 52.74\\
PSNR $\uparrow$ & \textbf{24.14} & 23.6 & 23.88 & 22.81 & 20.84 \\
SSIM $\uparrow$ & \textbf{0.795} & 0.786 & 0.794 & 0.784 & 0.662 \\
bitrate [kbps] $\downarrow$ & $\sim$5 & $\sim$3.1 & $\sim$8 & $\sim$3 & 10.6 \\
\end{tabular}
\end{center}
\caption{Average results of learned and classical metrics and also bitrate on test set. Arrows indicate if lower or bigger score mean better performance.}
\label{tab:lpips_etc}
\end{table}

The impact of the optimizations of the discriminator is given in Tab. \ref{tab:GAN_lambda}. The table shows the objective metrics depending on the amount ($\lambda$) of adversarial loss used and if gradient normalization (\textit{GN}) is present or not (\textit{no GN}).
Here it can be seen that the proposed discriminator optimizations allow for 4 times larger adversarial loss ultimately resulting in better quality. The last column shows the result from a training, where the discriminator became unstable, and the generated items were mostly noise.
\begin{table}[h!]
\begin{center}
\begin{tabular}{ r | l l l l}
        & $\lambda$ 4 GN & $\lambda$ 2 GN & $\lambda$ 1 no GN & $\lambda$ 4 no GN\\ 
\hline
LPIPS $\downarrow$ & \textbf{0.1793} & 0.1956 &  0.1993 & 0.721 \\ 
DISTS $\downarrow$ & \textbf{0.0998} & 0.109  &  0.1171 & 0.527 \\  
\end{tabular}
\end{center}
\caption{Impact of the optimizations of the discriminator. $\lambda$-values give the amount of adversarial loss used.}
\label{tab:GAN_lambda}
\end{table}

Finally we provide numbers of parameters and estimates of complexity of the presented model and of OSFV in Tab. \ref{tab:CMPLXTY}. The complexity is given in giga multiply-adds [MACs] per video frame. The proposed optimizations have only a very small impact on the complexity and memory size of FOMM. 
\begin{table}[h!]
    \begin{center}
    \begin{tabular}{ r | l l l l}
            & Nr of parameters & complexity [giga MACs] \\ 
    \hline
    FOMM & \num{59.79e6} & 54.96 \\ 
    ours & \num{59.87e6} & 55.22 \\ 
    ours no JAC & \num{59.72e6} & 54.73 \\ 
    OSFV & \num{173.2e6} & 483.85 \\  
    \end{tabular}
    \end{center}
    \caption{Numbers of parameters and estimates of complexity of the presented model and of OSFV.}
    \label{tab:CMPLXTY}
\end{table}

\section{Summary}
We present optimizations of a promising video codec that can satisfy the ever-growing hunger for video-communication data rate. 
Besides operating at very low bitrates this codec also runs at moderate computational complexity and with low delay. 
Our work shows that the Jacobian in FOMM may not be needed as full 2x2 matrices. Furthermore, we show that stabilizing the discriminator can further improve the quality. This stabilization is not limited to the proposed system and can be used in all learning based communication codecs.


\pagebreak
\bibliographystyle{IEEEbib}
\bibliography{refs}

\end{document}